\documentclass[%
 aip,
amsmath,amssymb,
preprint,%
numerical,%
]{revtex4-1}

\usepackage{graphicx}
\usepackage{dcolumn}
\usepackage{bm}
\usepackage{natbib}

\begin{document}


\title{Gas temperature dependent sticking of hydrogen\\on cold amorphous water ice surfaces of interstellar interest}

\author{E. Matar}
\email{Elie.Matar@u-cergy.fr}
 \altaffiliation[Present address: ]{Universit\'e Paris-Sud, ISMO, FRE 3363 du CNRS, Bat.351, 91405 Orsay, France}

\affiliation{LERMA-LAMAp, Universit\'e de Cergy-Pontoise, Observatoire de Paris, ENS, UPMC, UMR 8112 du CNRS; 5 mail Gay Lussac, 95000 Cergy Pontoise Cedex, France}

\author{H. Bergeron}
\email{herve.bergeron@u-psud.fr}
\affiliation{Universit\'e Paris-Sud, ISMO, FRE 3363 du CNRS, Bat.351, 91405 Orsay, France}

\author{F. Dulieu} 
\affiliation{LERMA-LAMAp, Universit\'e de Cergy-Pontoise  \& Observatoire de Paris, ENS, UPMC, UMR 8112 du CNRS; 5 mail Gay Lussac, 95000 Cergy Pontoise Cedex, France}
\author{H. Chaabouni}
\affiliation{LERMA-LAMAp, Universit\'e de Cergy-Pontoise  \& Observatoire de Paris, ENS, UPMC, UMR 8112 du CNRS; 5 mail Gay Lussac, 95000 Cergy Pontoise Cedex, France}
\author{M. Accolla}
\altaffiliation[Present address: ]{Universit\`{a} di Catania, DMFCI, viale Doria 6, 95125 Catania, Sicily, Italy}
\affiliation{LERMA-LAMAp, Universit\'e de Cergy-Pontoise  \& Observatoire de Paris, ENS, UPMC, UMR 8112 du CNRS; 5 mail Gay Lussac, 95000 Cergy Pontoise Cedex, France}
\author{J. L. Lemaire}
\affiliation{LERMA-LAMAp, Universit\'e de Cergy-Pontoise  \& Observatoire de Paris, ENS, UPMC, UMR 8112 du CNRS; 5 mail Gay Lussac, 95000 Cergy Pontoise Cedex, France}

\date{\today}

\begin{abstract}
Using the King \& Wells method, we present experimental data on the dependence of the sticking of molecular hydrogen and deuterium on the beam temperature onto non-porous amorphous solid water (ASW) ice surfaces of interstellar interest. A statistical model that explains the isotopic effect and the beam temperature behavior of our data is proposed. This model gives an understanding of the discrepancy between all known experimental results on the sticking of molecular hydrogen. Moreover it is able to fit the theoretical results of V. Buch {\it{et al.}} [Astrophys. J. (1991), {\bf 379}, 647] on atomic hydrogen and deuterium. For astrophysical applications, an analytical formula for the sticking coefficients of H, D, H$_2$, D$_2$ and HD in the case of a gas phase at thermal equilibrium is also provided at the end of the article.
\end{abstract}

\maketitle

\section{\label{sec:intro}Introduction}

Hydrogen is the most abundant element in the Universe. It constitutes 75\% of the total matter by mass and over 90\% by number of atoms. In dense clouds of the interstellar medium (ISM), hydrogen exists predominantly in its molecular form and is the chief constituent. The hydrogen molecule is of fundamental importance in the evolution of the Universe for two reasons. (a) Because of its high efficiency as a coolant it increases the rate of collapse of low mass interstellar clouds. (b) Once ionized by UV photons or by cosmic rays, it is of paramount importance in all the reaction schemes that form most of the molecular species in the gas phase.\cite{Dalgarno2000}
Moreover the hydrogen molecule enhances the sticking probability of hydrogen atoms and molecules on the icy interstellar dust grains.\cite{govers1980,schutte1976}

We know by now that gas-grain reaction are the most efficient route of molecular hydrogen formation in the ISM.\cite{watson1975,Duley1986} Two main mechanisms contribute to this formation route: (1) the Eley-Rideal (ER) mechanism in which a gas-phase hydrogen atom reacts directly with an adsorbed atom on the grain, and (2) the Langmuir-Hinshelwood (LH) mechanism in which both atoms are adsorbed and at least one of them diffuses on the surface of the grain to find its partner and to form a molecule. \cite{duley1996} Then the sticking of atomic and molecular hydrogen onto the interstellar dust grains plays a major role in understanding the surface chemistry in the ISM. 

In the dense interstellar medium dust grains (carbonaceous or silicates) are covered with icy mantles mainly composed of water. It is widely accepted that these mantles have an amorphous structure,\cite{hagen1983,Mathis1986,Tielens1987} and are covered with molecular hydrogen (1.5 -- 2.0~10$^{14}$~molecules.cm$^{-2}$).\cite{govers1980,buch1994,Congiu2009}

The sticking of atomic and molecular hydrogen on amorphous ice surfaces under interstellar conditions has been extensively studied theoretically over the years, but few experimental works have been carried out on the subject so far. The first theoretical studies were those of Hollenbach \& Salpeter,\cite{hollenbach1970} followed by Burke \& Hollenbach,\cite{burke1983} and Leitch-Devlin \& Williams.\cite{leitch-devlin1985} But the first intensive calculations were those of Buch \& Zhang.\cite{buch1991} These authors studied the sticking of H and D atoms on an amorphous water ice cluster made of 115 H$_2$O molecules by using classical molecular dynamics (CMD) simulations. They calculated the sticking coefficients $S(E)$ for several beam kinetic energies $E$ ranging from 50~K to 600~K. The fitting of their results was done using the simple exponential decay function $S(E)$=$e^{-E/E_0}$, where $E$ is the kinetic energy of the incident gas and where $E_0$=200~K for D atoms and $E_0$=102~K for H atoms ($\sim$2 times lower than that of D). 

Matsuda {\it et al.},\cite{masuda1998} and Takahashi {\it et al.},\cite{takahashi99} also studied the sticking of atomic hydrogen on amorphous ice slab made of 1000 H$_2$O molecules (40\AA$\times$40\AA$\times$20\AA) at 10~K using CMD calculations. They calculated the sticking probability of hydrogen atoms as a function of the incident beam temperature (i.e., the temperature of the gas in the beam) and found that, for a kinetic energy $E_i$=10~K, the sticking probability of a hydrogen atom is unity. 

In the same context, classical trajectory (CT) calculations were performed by Al-Halabi \& van Dishoeck.\cite{Al-Halabi2007} Their results on the sticking probability of H atoms on amorphous solid water (ASW) ice (6 bilayers of 360 H$_2$O molecules) at 10~K were fitted by the same decaying exponential function used by Buch \& Zhang.\cite{buch1991} These authors found that $S(E)=\alpha e^{-E/E_0}$, where $E$ is the kinetic energy of the incident atoms, $E_0$=300~K (as compared with 102~K for Buch) and $\alpha=1$ is a constant parameter. With these fitting function and parameters, the sticking coefficient of an H atom with $E$=10~K on a surface at 10~K is equal to $0.97$ and with $E$=300~K it is equal to $0.37$.

To date, only one set of experiments on the sticking of atomic hydrogen on amorphous water ice surfaces can be found in the literature,\cite{schutte1976} and few experimental works have been conducted to measure the sticking coefficients of hydrogen and deuterium molecules on the same ice surfaces at low temperature.\cite{govers1980,Hornekaer2003}

Schutte {\it et al.},\cite{schutte1976} and Govers {\it et al.},\cite{govers1980} presented results on the variation of the sticking and accommodation of molecular hydrogen (deuterium) with surface coverage by H$_2$ (D$_2$). They both used bolometer experiments to study the sticking coefficient on a surface of a cryodeposit of H$_2$O, N$_2$ and Ar in the 3.5~K--15.5~K range. They found that at a surface temperature $\leq$10~K, the sticking coefficient of impinging H$_2$ molecules on an initially H$_2$ free surface increases very slightly, when the surface temperature decreases. This means that the sticking coefficient is rather independent of the surface temperature in the 3.7~K--10~K range. These authors also found that the sticking probability is highly dependent on the H$_2$ coverage of the surface, and that it increases with the increasing amount of adsorbed H$_2$. Govers {\it et al.},\cite{govers1980} found that the sticking coefficient of H$_2$ is equal to $0.08\pm0.05$ and that of D$_2$ equal to $0.27\pm0.05$ for beams at room temperature and with an incidence angle of $45^{\circ}$ with respect to the normal to the surface. [These values of the sticking coefficients are obtained directly from their figures 5 and 6. They are not exactly those indicated in the abstract of their article.]

Hornekaer {\it et al.},\cite{Hornekaer2003} studied the HD formation efficiency on both porous ($p$-) and non-porous ($np$-) ASW ice at 8~K using the TPD (Temperature Programmed Desorption) technique. These authors measured the sticking coefficient of D$_2$ with the King \& Wells method,\cite{King1972} shortly described in section II. They found $S_{D_2}$=0.20$\pm$0.15 for a D$_2$ beam direction perpendicular to the surface, at room temperature, on an $np$-ASW ice at 8~K. 

Amiaud {\it et al.},\cite{amiaud2007} used the same method also for D$_2$ and found $S_{D_2}$=0.38$\pm$0.05 for a D$_2$ beam at room temperature and with an incidence angle of 43$^{\circ}$ with respect to the normal. 

In the present article, using the King \& Wells method, we give the first experimental results on the variation of the sticking coefficient of molecular hydrogen and deuterium on ASW ice surfaces as a function of the incident molecular effusive beam temperature. 

Moreover, in order to explain: (a) the isotopic effect and the beam temperature behavior of our data, (b) the discrepancy between all known experimental results (including our data) on the sticking coefficients of H$_2$ and D$_2$, a simple physical model involving few parameters is developed. To our knowledge, it is the first time that a model reproducing a so wide range of results is proposed. This model is divided in two parts. The first part is devoted to the description of the physical process of sticking alone, while the second part describes the effect of the effusive molecular beam velocity distribution on the measurements. 

The article is organized as follows. In section II we briefly describe the experimental setup and procedures. In section III we present our experimental results. In section IV we develop our model. In section V we apply our model to analyze our data and those of previous experiments (also based on molecular beams). We then discuss the obtained results (Sec. V) before concluding (Sec. VI). Finally the strength of our model is tested (Appendix A) by fitting the theoretical results of V. Buch {\it et al.} \cite{buch1991} on atomic hydrogen and deuterium.

\section{\label{sec:Exp}Experimental}
\subsection{\label{sec:setup}Apparatus}
The experiments are performed using our FORMOLISM (FORmation of MOLecules in the InterStellar Medium) set-up, that we will briefly describe here (more details can be found in Amiaud {\it et al.}\cite{amiaud2007} FORMOLISM is an apparatus devoted to study the reaction and interaction of various atomic and molecular beams on dust grain analogs under interstellar conditions. 

\begin{figure}
\includegraphics[scale=0.5]{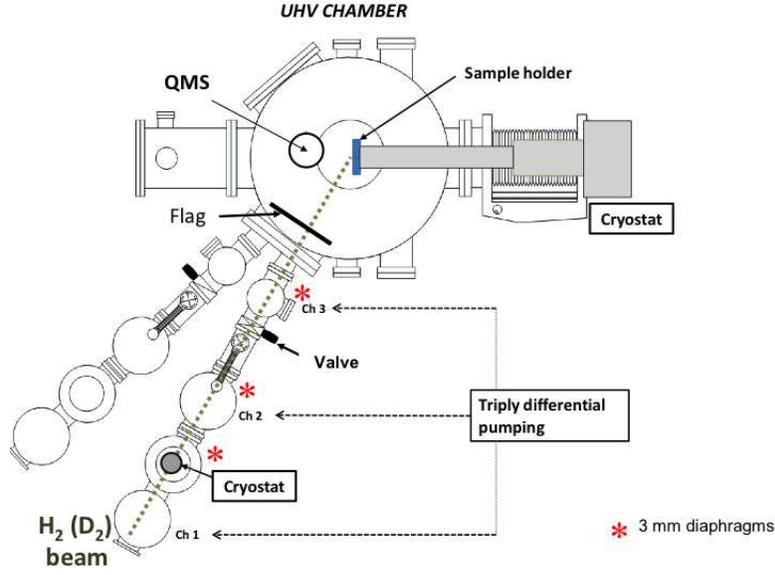}
\caption{Experimental setup} 
\label{setup}
\end{figure}

This apparatus (figure \ref{setup}) is composed of an ultra-high vacuum (UHV) chamber with a base pressure of $\sim$0.5~10$^{-11}$ mbar, of an oxygen-free high-conductivity copper sample holder (1 cm in diameter), in good thermal contact with a cold finger of a He closed-cycle cryostat (ARS-Displex DE-204S), that can be cooled down to 8 K, and of a translatable quadrupole mass spectrometer (QMS Hiden HAL-3F) that is usually used in two positions, in front of or above the sample holder.

The temperature is measured using a calibrated silicon diode clamped on the sample holder and connected to a temperature controller (Lakeshore 340). The temperature can be controlled to $\pm0.2$ K with an accuracy of $\pm1$ K in the range of $8-400$~K. A micro-channel array doser (1 cm in diameter) is moved in front of the sample to expose it to H$_2$O vapor and to grow ASW ice films. The procedure for $np$-ASW ice film growth is already described under procedures (sample temperature).

Hydrogen and/or deuterium reactants are introduced into the UHV chamber via two separated triply differentially pumped beam lines aimed at the sample holder. For this experiment we only use the first beam line, making a 62$^{\circ}$ angle with the normal to the surface, where the gas flows through an aluminum nozzle connected to a He closed-cycle cryostat, in order to cool it down to a controlled temperature $T_B$ before entering in the UHV chamber. A valve is located between the second and the third stage of the beam-line (separated by a 3~mm diaphragm) that is used in these experiments to create an effusive beam. In fact, opening the valve between the two stages allows us to estimate the amount of molecules that are diffusing from one chamber to the other out of the beam. At the entrance of the main chamber, also separated from the third stage by a 3~mm aperture, there is a flag that we use to intercept the beam, prohibiting the species to directly reach the surface of the sample holder. 

The effusive character of the beam at the end of the line (entry in the UHV chamber) is deduced from the Knudsen criterion that allows to distinguish effusive beams from free beams (supersonic) in the first chamber, but no direct measurement of the velocity distribution has been done. 

We call beam temperature ($T_B$) the temperature of the gas phase in the last part of the injection line (made ofÊ aluminum) : $T_B$ is our tunable parameter. Let us notice for what follows that an effusive beam is not a gas phase at equilibrium (in the sense of thermodynamics) and then its temperature is not defined. Then $T_B$ is a parameter (having the dimension of a temperature) that characterizes the velocity distribution of the beam in the UHV chamber, but the convenient expression ``beam temperature" is abusive (or misleading).

\begin{figure}
\resizebox{\hsize}{!}{\includegraphics{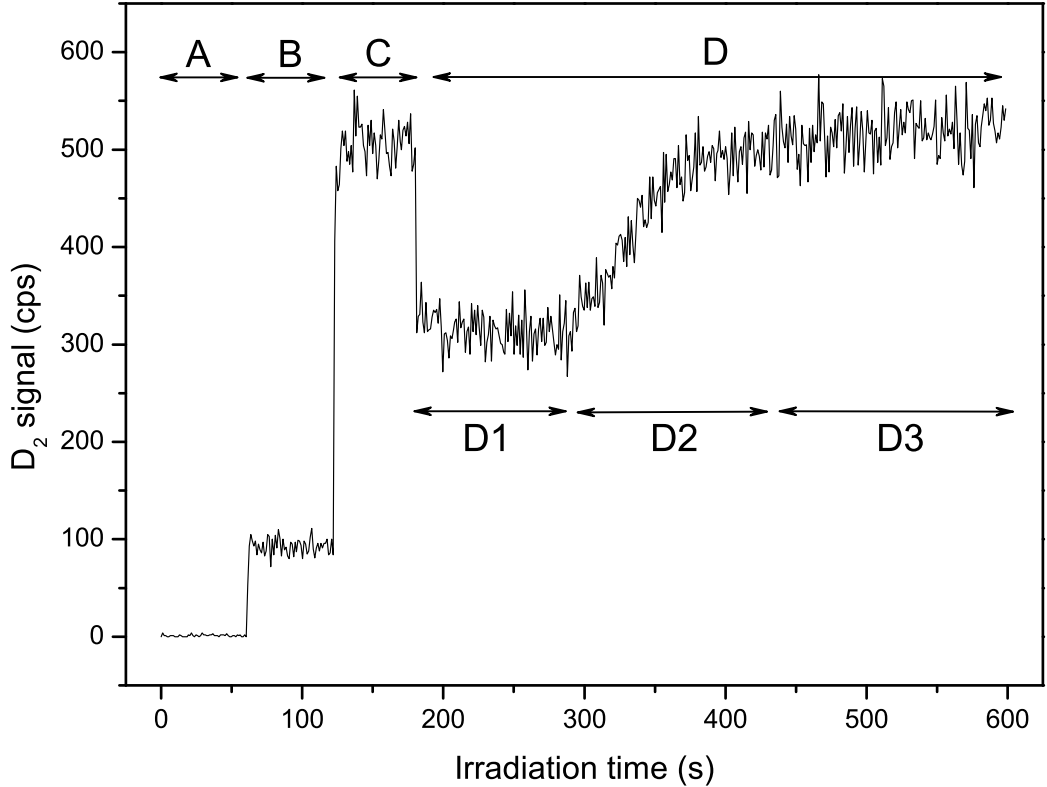}}
\caption{D$_2$ signal registered with the QMS during the 4 steps of each experiment (See text for detailed explanation). (A: Valve closed, B: Valve slightly open+flag, C: Valve widely open+flag, D: Beam directly aimed at the surface). The beam temperature $T_{B}$ is 240~K and the sample at 10~K.} 
\label{typical}
\end{figure}

\subsection{\label{sec:proc}Procedures}
In order to perform our experiments, we first move the micro-channel array doser 20 mm away in front of the sample holder maintained at 120 K. We then introduce H$_2$O vapor by maintaining the pressure in the UHV chamber at 2~10$^{-9}$ mbar. With this method we grow $np$-ASW ice \cite{kimmel2001a,kimmel2001b, fillion2009}. Previous calibrations showed that we need $\sim$5~minutes to grow 100~ML (1.0 ML=10$^{15}$ cm$^{-2}$) of $np$-ASW. Finally we cool down the surface temperature $T_{S}$ to 10 K when the partial pressure of water monitored by the QMS is stable and close again to the base pressure in the UHV chamber.

The sticking probability of reactive molecules on a surface is usually measured by the well-established beam reflectivity technique of King and Wells.\cite{King1972} In this technique, the intensity of the molecules scattered from the surface, recorded by the QMS in the UHV chamber, is used as a direct measurement of the sticking coefficient. For this we procede as follows: 

After the preparation of a stable and pure D$_2$ (H$_2$) beam, the $np$-ASW ice surface, held at 10 K, is then exposed to the D$_2$ (H$_2$) beam at a chosen temperature.
The scheme of a typical experiment is shown in figure \ref{typical}. It is divided into four steps during which the QMS is above the surface and then monitoring the indirect D$_2$ (H$_2$) signal reflected from the surface. (A) We first start monitoring the signal for 60 seconds without introducing any D$_2$ (H$_2$) into the UHV chamber. (B) Secondly we 
introduce D$_2$ (H$_2$) into the chamber by slightly opening  the valve (between stages 2 and 3)
but still blocking the beam and adding the flag in order to create a constant background pressure. This will allow us to estimate the amount of molecules that are diffusing out of the beam. Intercepting the beam with the flag ensures that no direct molecules are reaching the $np$-ASW ice surface. (C) In the third step, the flag still intercepting the beam, we fully open the valve completely for 60 additional seconds to obtain a full indirect flux of D$_2$ (H$_2$). (D) In the last step the flag is completely removed enabling the molecules to directly reach the surface. This fourth step may span up to 600 seconds and is decomposed in three stages. In the first stage (D1), the QMS signal is linearly decreasing for $\sim100$ seconds and then it starts increasing rapidly (D2) before it stablizes after $\sim100$ seconds (D3). The same experiment is then repeated for several beam temperatures of D$_2$ (H$_2$), ranging from 28 to 350 K.

After the first $\sim$100~s, the signal starts to rise rapidly (D2) corresponding to a decrease in the sticking coefficient. This signal rise is obviously due to molecules that begin to desorb from the surface because their residence time becomes close to the time between two arrivals of impinging molecules.\cite{amiaud2007} The plateau at the end of this step (D3) corresponds to a steady state regime where the number of sticking molecules is equal to the number of desorbing molecules.

\begin{figure}
\resizebox{\hsize}{!}{\includegraphics[scale=0.5]{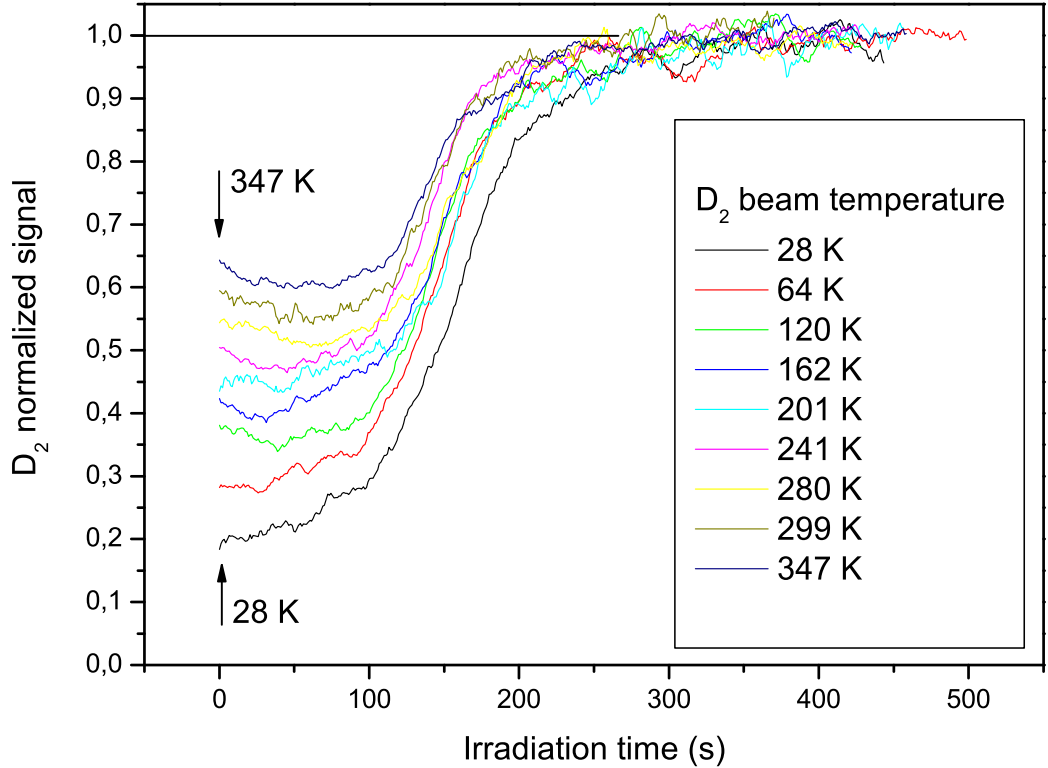}}
\caption{Sticking of D$_2$: normalized D$_2$ QMS signal during step D of figure \ref{typical} for several D$_2$ beam temperatures, ranging from 28 to 347 K. It shows how the sticking signal (stage D1), of impinging molecules, varies with beam temperature $T_{B}$. The signal is normalized to its final steady value. } \label{Dstick}
\end{figure}

\section{\label{sec:results}Results}
This typical experiment described above is known as the King and Wells method \cite{King1972}. It encompasses the four steps A to D. As in our experiments, the value of the D$_2$  signal in step (D3) is comparable to that of step (C)  with the flag in the beam (see figure \ref{typical}) we have started all measurements at the fourth step. Then D$_2$ (H$_2$) irradiation begins when the flag is removed and the beam then directly aimed at the surface.
Experimental results are shown in figure~\ref{Dstick} for the deuterium beam.

Figure~\ref{Dstick} shows the D$_2$  normalized signal as function of the irradiation time for different D$_2$  beam temperature, starting from the fourth step (D1) when the flag is removed. 

After the removal of the flag (step D starting here at t=0), the signal drops down dramatically compared to step C. This signal lowering is explained by the sticking of the D$_2$ molecules on the ice surface. One can also see that during the first stage (D1), the behavior of the D$_2$ signal is highly dependent on the beam temperature $T_{B}$.

A first remark concerns the initial value of the D$_2$ signal. One can clearly see that the lower the temperature is the lower the signal, then increasing with increasing $T_{B}$. This could be explained by the fact that for low $T_{B}$, the kinetic energy of the molecules is low, making collisions with the ice surface less elastic, thus increasing the probability for a molecule to stick onto the surface. As a result, the sticking coefficient of D$_2$ increases when $T_{B}$ decreases.
A second remark is that, for high $T_{B}$, the signal starts with a noticeable linear decrease before it reaches the second stage D2. This almost linear decrease disappears at lower temperatures. This decrease has been explained by Govers {\it et al.}.\cite{govers1980} These authors interpreted this decrease in the signal as an increase in the sticking coefficient induced by the presence of molecules already adsorbed on the surface. In fact, when a gas-phase D$_2$ (H$_2$) molecule impinges on a D$_2$ (H$_2$) molecule already adsorbed on the ice surface, the accomodation is greatly enhanced,\cite{schutte1976} thus enhancing the sticking coefficient of the impinging molecule. On the other hand, the disappearance of this linear decrease for lower $T_{B}$ might be explained by the fact that at very low temperatures the sticking coefficient is already at its maximum and cannot increase any further. Then the signal rises slowly right from the start before reaching the rapid increase of the second stage.

An absolute sticking coefficient can be derived from the curves in figure~\ref{Dstick}. The measured yield of D$_2$ (H$_2$) molecules $Y(t)$ is the sum of molecules reflected by the surface $R_{f}(t)$ (figure \ref{typical}, part D1) and a constant background factor $B$ (figure \ref{typical}, part B): 
\begin{equation}
Y(t)=B+R_{f}(t).
\end{equation}
The sticking coefficient $S(t)$ is equal to the ratio between the non-reflected part of the signal and the incoming flux $F$: 
\begin{equation}
S(t)=\frac{F-R_{f}(t)}{F}.
\end{equation}
In the steady state regime (figure \ref{typical}, part D3), where $t=\infty$, the incoming flux is equal to the reflected molecules. In this case $F=R_{f}(\infty)$ and then we can derive the sticking coefficient\cite{amiaud2007}: 
\begin{equation}
S(t)=\frac{Y(\infty)-Y(t)}{Y(\infty)-B}.
\end{equation}

\begin{figure}
\resizebox{\hsize}{!}{\includegraphics{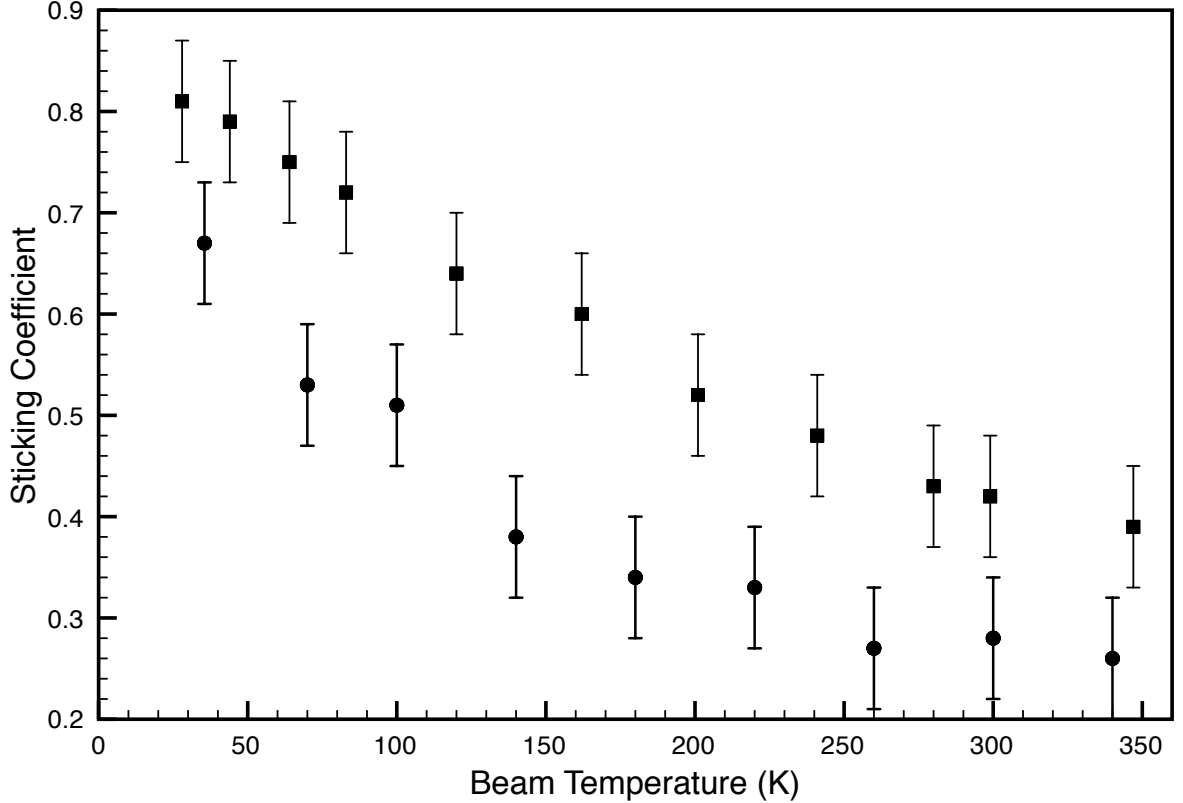}}
\caption{Experimental sticking coefficient for H$_2$ and D$_2$ ranging from $T_B=28 \text{ K}$ to $347 \text{ K}$. Plain circles are for H$_2$ and squares for D$_2$. The absolute uncertainties are equal to $\pm 0.06$ and are calculated by measuring the dispersion of the same sticking coefficient measured several times. } \label{ExpResults}
\end{figure}

With this method we calculate the absolute sticking coefficients of H$_{2}$ and D$_{2}$ for different beam (gas) temperatures $T_B$ as shown in figure~\ref{ExpResults} (the absolute sticking coefficient is defined as $S(t=0)$).


\section{\label{sec:model} Model}
In this section we present a statistical model that explains and fits these experimental results as well as some previously published experimental (from Govers {\it et al.},\cite{govers1980} Hornekaer {\it et al.},\cite{Hornekaer2003} Amiaud {\it et al.} \cite{amiaud2007}) and theoretical (from Buch \& Zhang \cite{buch1991}) results on sticking probabilities of H$_2$, D$_2$, H and D.

\subsection{\label{subsec:framework} Framework and assumptions}
In order to model the experiment, we consider an amorphous surface, flat on large scale (with respect to atomic scale), at a temperature $T_{S}$. In the following $T_S$ is assumed  to be fixed at the temperature of our experiment ($T_S \simeq 10 \text{ K}$) and we omit this parameter in all functions to simplify notations. 

We also have a gas phase (in the half-space above the surface) composed of particles of mass $m$ coming from an effusive beam that irradiates the surface. The gas phase velocity distribution is assumed to be that of an effusive beam (and not that of a thermal equilibrium):  this distribution and some angle-dependent effects explain the difference between our results and already published experimental results. 

The heart of the model is the representation of the sticking process for which we consider that the sticking of the particles is essentially due to physisorption. 

First, if the surface was ordered (not amorphous) then we could assume that each impinging particle has only one possibility to stick on the surface: if it does not stick immediately when it hits the surface then it is lost. Moreover in this situation of an ordered surface, the potential of gas-surface interaction essentially depends on the coordinate normal to the surface. Thus the main effect on sticking to the surface is due to the normal component $v_{\bot}$ of the gas particle velocity.

In our case the surface is amorphous. This is treated as follows. Since we are interested in probabilities and not in the details of trajectories, we can always define the probability of sticking $S_P$ for a particle (coming from the gas phase) that hits the surface (for the first time) at a given point $P$, i.e. the probability that the resulting trajectory leads to a stuck particle. If the surface is very disordered, the probability of sticking at the point $P$ and at the point $P'$ very close to $P$ can be very different, removing any macroscopic physical meaning to $S_P$. So we imagine some coarse graining of the surface that divides it into a family of cells $\{ \mathcal{C} \}$ such that each cell $\mathcal{C}$ can be represented by averaged local properties (for example an averaged normal vector $\vec{n}_{\mathcal{C}}$). We define also the probability of sticking $S_{\mathcal{C}}$ for a particle hitting the surface into the cell $\mathcal{C}$. Of course cells are sufficiently small to have different physical properties from place to place (in particular sticking properties) and this represents (statistically speaking) the disordered nature of our surface on large scale.

Under these conditions, we can roughly transform the sticking dependence on normal velocity $v_\bot$ of the ordered situation, into a dependence of $S_{\mathcal{C}}$ on the modulus $v$ of the velocity. This change is justified by the random topological nature of the surface on the cell scale: the macroscopic normal has no physical meaning on the cell scale and then the component $v_\bot$ is not meaningful. \\
Now we assume that each cell $\mathcal{C}$  is characterized by a phenomenological velocity $c(m,\mathcal{C})$ (depending {\it{a priori}} on the mass of the impinging particles), such that if the modulus $v$ of the impinging particle velocity verifies $v> c(m,\mathcal{C})$, the particle rebounds in the gas phase, and if $v < c(m,\mathcal{C})$ the collision is sufficiently inelastic and the particle is stuck. Since our surface is amorphous, $c(m,\mathcal{C})$ is randomly different from cell to cell . Therefore we will have to introduce a probabilistic distribution that represents the different values of $c(m,\mathcal{C})$.\\
In general the phenomenological velocity $c(m,\mathcal{C})$ must not only depend on the mass of the impinging particle, but also on the particle-surface potential interaction. However in our case we are interested in the molecules H$_2$ and D$_2$ that have the same electronic properties, so that the particle-surface interaction must be essentially the same.

\subsection{\label{subsec:stickSvz} The sticking process (coefficient $S(G, v)$)}
$S(G, v)$ is defined as the sticking probability of particles of given species (G) and given velocity.
This quantity can be put in correspondence with the coefficient usually called $S(E)$ (Buch \& Zhang,\cite{buch1991} Al-Halabi \& van Dishoeck \cite{Al-Halabi2007}) which is the sticking probability for a given kinetic energy. 

Two probabilities contribute to this sticking probability. To simplify, we consider that these contributions are independent and we multiply them to obtain the final sticking probability $S(G, v)$.

\begin{itemize}

\item The first contribution $S_{0}(G)$ is taken to be a characteristics of the species G (independent of the velocity). Focusing differently, this quantity can be roughly seen as the probability  for a given particle of vanishing kinetic energy ``put" on the surface (on a random place) to stick on it. 

\item The second contribution $P(v)$ is the probability to have $v<c(m,\mathcal{C})$. It thus involves the probability distribution $g(c)$ corresponding to the values of $c(m,\mathcal{C})$. This latter is assumed to have the form $g(c)= \phi(c/c_0(m))/c_0(m)$ where $\phi$ is a (unknown) surface-dependent function and $c_0(m)$ is a parameter having the dimension of a velocity. The function $\phi$ must verify the condition $\int_0^\infty \phi(x)dx=1$ (probability normalization).

Thus $P(v)$ writes
\begin{equation}
P(v)=\Phi \left( \frac{v}{c_0(m)} \right)  \text{ with }   \Phi(a)=\int_a^\infty \phi(x)dx.
\end{equation}
\end{itemize}

\noindent Finally, the sticking probability $S(G,v)$ can be written as: 
\begin{equation}
S(G,v)=S_{0}(G)\Phi \left( \frac{v}{c_0(m)} \right).
 \label{probastickv}
\end{equation}
Even if we do not know the function $\Phi$, we can say from its definition that $\Phi$ is a positive decreasing function such that $\Phi(0)=1$ and $\Phi(\infty)=0$.
 
It is noteworthy that the function $\Phi$  does not depend on the molecule but only on the surface. The only parameters depending on the molecule being $S{_0}( G)$ and $c_0(m)$.

In the case of H$_2$ and D$_2$, the experimental data seem to show a negligible dependence of $c_0$ on the mass. {\it So in what follows the mass dependence of $c_0$ is neglected}.

\subsection{\label{subsec:stickST} The measured sticking coefficient $S(G,T_B)$}
Actually, in view of the experimental results, one has to go one step further, and get from $S(G,v)$ the sticking probability $S(G,T_B)$ as a function of $T_B$ (the so-called beam temperature). This requires the introduction of a velocity distribution function $f(v,T_B)$. The previous formula (Eq. \ref{probastickv}) must be changed into
\begin{equation}
\label{eqn:stickwithtemp}
S(G,T_{B})=\int^{\infty}_{0}S(G,v)f(v, T_{B})dv.
\end{equation}

Actually $f(v,T_B)$ is not merely the Boltzmann distribution. Indeed, for a surface being irradiated with a direct effusive beam, a privileged direction exists and the real velocity distribution of the effusive beam must be taken into account. The velocity distribution of an effusive beam \cite{Dunoyer} $f_b(\vec{v},T_B)$  is 
\begin{equation}
f_b(\vec{v}, T_{B})=Z^{-1}v_{b} e^{-\frac{m \vec{v}^{2}}{2k_{B}T_{B}}}
\end{equation}
where $v_b=\vec{v}.\vec{u}_b>0$ is the velocity component in the beam direction specified by the unit vector $\vec{u}_b$ and $Z$ is a normalization constant. {\it In that case the gas phase is not at thermal equilibrium, and $T_B$ is not the gas phase temperature, since this quantity is in fact undefined}.

If we call $\theta$ the angle between the beam direction and the normal to the surface of our sample, two extreme situations can occur. In the first case $\theta \simeq 90^\circ$ (the exact value $90^\circ$ is of course not interesting in practice: molecules cannot hit the surface). Due to the geometry of the experiment, the surface can only see the {\it central part of the effusive beam}, that is the one-dimensional velocity distribution in the beam direction $\vec{u}_b$ and thus $v_b=v$. In the second case $\theta=0^\circ$. In that situation we can assume that the surface sees the {\it full effusive beam} velocity distribution. But we are only interested in the velocity modulus. So integration over spherical coordinates \cite{Dunoyer} leads to the replacement $v_b \to v^3$ (the $v^2$ supplementary dependence arises from the volume element). Thus the angle dependent velocity modulus distribution $f(\theta,v,T_B)$ seen by the sample writes
\begin{equation}
\label{eqn:expdistrib}
f(\theta,v,T_B)=Z^{-1} v^{\alpha(\theta)} e^{-\frac{m v^2}{2 k_B T_B}},
\end{equation}
where we choose $\alpha(\theta)=1+2 \cos^2(\theta)$ to interpolate between the two extreme cases just discussed [because of the symmetry with respect to the sample plane, no linear term in $\cos \theta$ is involved ]. $Z$ is a normalization constant obtained from the constraint $\int_0^\infty f(\theta,v,T_B) dv=1$. 

From Eqs \ref{probastickv}, \ref{eqn:stickwithtemp} and \ref{eqn:expdistrib}, we thus obtain after some calculations
\begin{eqnarray}
\label{eqn:sintform}
S(G, \theta, T_{B}) &&=\frac{S_{0}(G)}{\Gamma \left( \frac{\alpha(\theta)+1}{2} \right)} \\
&&\times \int^{\infty}_{0} u^{\frac{\alpha(\theta)-1}{2}} e^{-u}\Phi \left( \sqrt{\frac{T_{B}u}{T_{0}(m)}} \right)du,\nonumber
\end{eqnarray}
where $T_{0}(m)=c_0^{2} m/(2k_{B})$ and $\Gamma$ is the Euler function. In our case $\theta=62^\circ$ and $\alpha(\theta) \simeq 1.44$.

\subsection{\label{subsec:MassandTemp} Mass and temperature dependence of the sticking coefficients}
To simplify notations, we replace $T_B$ by $T $, $S(G,\theta,T_B)$ by $S_{H_2}(\theta,T)$ or $S_{D_2}(\theta,T)$, and $S_{0}(G)$ by $S_{0}(H_2)$ or $S_{0}(D_2)$ to represent the sticking coefficients of H$_{2}$ and D$_{2}$ respectively. We also introduce the masses $m_{H_2}$ and $m_{D_2}=2m_{H_2}$, we then have $T_{0}(H_2)$ and $T_{0}(D_2)=2T_{0}(H_2)$ due to proportionnality to the mass.
If we define the mathematical function $\hat{\Phi}$ as:
\begin{equation}
\hat{\Phi}(\theta, x)=\frac{1}{\Gamma \left( \frac{\alpha(\theta)+1}{2} \right)} \int^{\infty}_{0} u^{\frac{\alpha(\theta)-1}{2}} e^{-u}\Phi(\sqrt{xu})du,
\label{eqn:phithetax}
\end{equation}
we then have from Eq. \ref{eqn:sintform}
\begin{equation}
\left\{
\begin{array}{l}
S_{H_2}(\theta, T)=S_{0}(H_2)\hat{\Phi}(\theta, \frac{T}{T_{0}(H_2)})\\
S_{D_2}(\theta, T)=S_{0}(D_2)\hat{\Phi}(\theta, \frac{T}{T_{0}(D_2)})
\end{array},
\right.
\label{sdshformula}
\end{equation}
with $T_{0}(D_2)=2T_{0}(H_2)$. We can deduce finally:
\begin{equation}
S_{H_2}(\theta, T/2)=\frac{S_{0}(H_2)}{S_{0}(D_2)}S_{D_2}(\theta,T).
\label{sdshequivalence}
\end{equation}
This means that under our assumptions, the experimental data of $S_{D_2}(\theta,T)$ should be equivalent to that of $S_{H_2}(\theta,\frac{T}{2})$ up to a renormalization. In the remainder we call this scaling law a renormalization-dilation transform.

This result is very simple and very interesting because it is independent of the form of the functions $\Phi$ and $\alpha(\theta)$ that we use, and then it lends itself to a direct test of our theoretical assumptions. Then according to our model, the isotopic effect on our data is essentially explained by the factor $2$ in Eq. (\ref{sdshequivalence}) which is just the mass ratio between D$_2$ and H$_2$.

\section{Data analysis from our model}

\subsection{\label{subsec:CompwithExp} Our experiment}
\begin{figure}
\resizebox{\hsize}{!}{\includegraphics{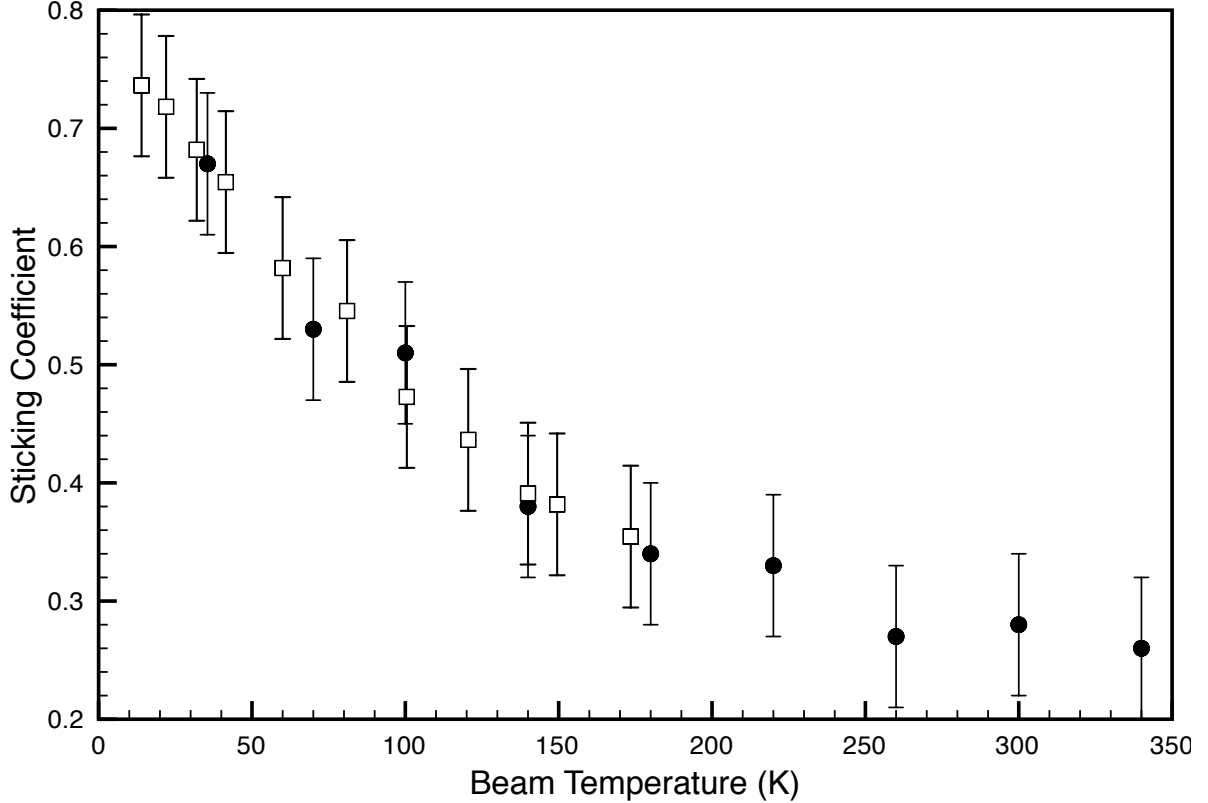}}
\caption{Sticking coefficients: test of our model after our renormalization-dilation tranform of Eq. \ref{sdshequivalence}. Closed circles: H$_{2}$ data, open squares: D$_{2}$ data transformed with Eq. \ref{sdshequivalence}.  These results are obtained with $S_0(D_2)/S_0(H_2)=1.1$.} \label{DH-ratio}
\end{figure}

We use Eq. (\ref{sdshequivalence}) to represent on the same graph our experimental results for H$_2$ and D$_2$ with the ratio $S_{0}(D_2)/S_{0}(H_2)=1.1$ (value obtained after optimization). The result is shown in figure \ref{DH-ratio}. This figure corroborates the renormalization-dilation transform in equation (\ref{sdshequivalence}). 

Of course to go further, in particular to obtain a fit of these experimental results, we need to fix the unknown function $\Phi$. In making this choice we referred to the work of refs. (\onlinecite{buch1991}), (\onlinecite{Al-Halabi2002}) and (\onlinecite{Al-Halabi2007}) which suggests $\Phi(x)=e^{-x^2}$.  The corresponding transformed function $\hat{\Phi}$ is 
\begin{equation}
\hat{\Phi}(\theta, x)=(1+x)^{-\frac{\alpha(\theta)+1}{2}}.
\end{equation}

\begin{figure}
\resizebox{\hsize}{!}{\includegraphics{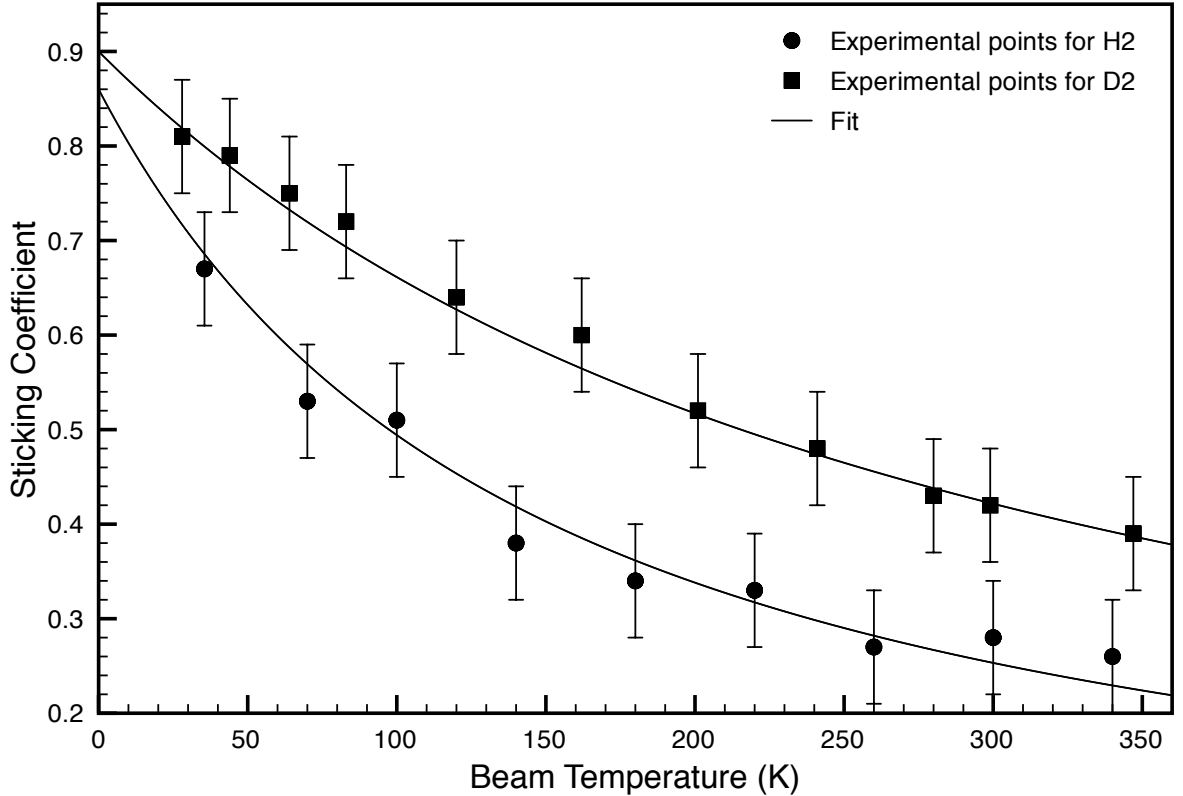}}
\caption{Sticking coefficients: fit of the experimental data obtained for $\Phi(x)=e^{-x^2}$, $S_{0}(H_2)=0.86$, $S_{0}(D_2)=0.9$ and $T_{0}(H_2)=174$ K. Closed circles represent H$_2$ and closed squares D$_2$ experimental points, respectively.} \label{firstfit}
\end{figure}

 The fits (Fig. \ref{firstfit}) are good for $T > 80 \text{ K}$, but the concavity of the curves (especially for D$_2$) in the domain $T < 80 \text{ K}$ does not seem to be the right one. 
 
 To improve this feature we have used the function $\Phi(x)=(1+x^2)e^{-x^2}$. When plugged in our model with adequate changes, it provides an excellent fit of the theoretical results of Buch \& Zhang \cite{buch1991} for H and D atoms (see Appendix A). This test is interesting because it involves only the function $S(G,v)$ of our model (and then the function $\Phi$) and not the angle-dependent velocity distribution.
 
The corresponding function $\hat{\Phi}$ is
\begin{equation}
\label{eqn:phicomplete}
\hat{\Phi}(\theta, x)=\frac{1+\beta(\theta) x}{(1+x)^{\beta(\theta)}},
\end{equation}
with $\beta(\theta)=0.5(\alpha(\theta)+3)=2+\cos^2(\theta)$. In our case $\beta(\theta=62^\circ)=2.22$.

\begin{figure}
\resizebox{\hsize}{!}{\includegraphics{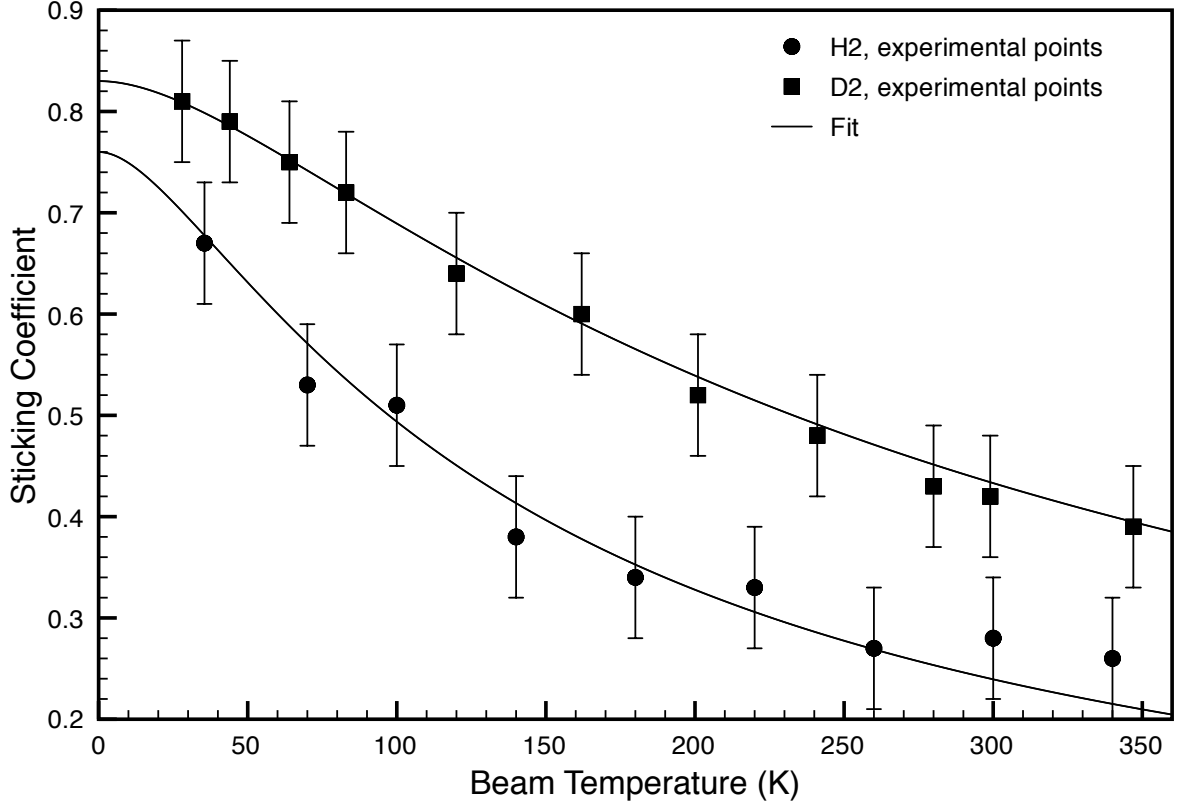}}
\caption{Fits of the experimental results for H$_2$ and D$_2$ obtained with the function $\Phi(x)=(1+x^2) \exp(-x^2)$. $S_0(H_2)=0.76$, $S_0(D_2)=0.83$ and $T_0(H_2)=87 \text{ K}$.} \label{figexpPlusBuch}
\end{figure}

Fig. \ref{figexpPlusBuch}  shows the fits of our data obtained with this new function: the fits are seen to be quite satisfactory in the whole $T$ range. We assume in what follows that the corresponding values $S_0(H_2)=0.76$, $S_0(D_2)=0.83$, $T_0(H_2)=87 \; \rm{K}$ and our function $\Phi$ are the physical parameters of the sticking process.

\subsection{\label{subsec:Testonexpresults} Comparison with other experimental results}

 The different experimental results (based on beam experiments) are presented in table \ref{tab:table1} where we add our results (for the same temperature). We see immediately their lack of compatibility if we decide to compare them directly. But the situation is changed if we compare them through our model, because these experiments were made with different values of the angle $\theta$ that modifies the velocity distribution seen by the sample. 
 
The result of Amiaud {\it et al.} \cite{amiaud2007} is expected to be well-reproduced by our angle-dependent distribution, because the experiment is based on exactly the same experimental set-up (except the angle) and the angular interpolating function $\alpha(\theta)$ has been partly elaborated for this purpose. We can be optimistic for the experimental result of Hornekaer {\it et al.},\cite{Hornekaer2003} corresponding to a perpendicular beam ($\theta=0$), because in that case we can assume that the sample sees the full distribution of the effusive beam, and this corresponds to one of the constraints imposed to our distribution. In the case of Govers {\it et al.},\cite{govers1980} the situation is different. The experiment was made with an angle $\theta=45^\circ$: some specific geometrical parameters of this experiment (different from our apparatus) are able to play a role in the angle-dependent velocity distribution seen by the sample.

 The comparison is presented in table \ref{tab:table1}. The values obtained confirm the above expectations.
  
 We may tentatively assume that the unique reason of the model breakdown for the Govers {\it et al.} data is due to a bad estimate of the velocity distribution parameter $\alpha$. From Eqs. \ref{sdshformula} and \ref{eqn:phicomplete}, we deduce that the sticking coefficient obtained by Govers {\it et al.} must verify
 \begin{equation}
 \label{eqn:stickbetatest}
 S(T)=S_0 \frac{1+\beta T/T_0}{(1+T/T_0)^\beta},
 \end{equation}
 where $S_0$ and $T_0$ are our values (for H$_2$ and D$_2$), but $\beta$ is now an unknown coefficient (identical for H$_2$ and D$_2$).
 
 Using the results of Govers {\it et al.} for H$_2$ and D$_2$, we can solve Eq. \ref{eqn:stickbetatest}, and we obtain two values of $\beta$ (one for H$_2$ and one for D$_2$). If our assumption is correct these values must be very close. 
 
 Taking into account the uncertainties on the sticking coefficients, we obtain $\beta(\rm{H}_2) \in [2.78, 4.01]$ and $\beta(\rm{D}_2) \in [2.70, 3.23]$. Now recalling that the full distribution of the effusive beam corresponds to $\beta=3$ which is a physical bound, we can restrict the previous intervals to $\beta(\rm{H}_2) \in [2.78, 3]$ and $\beta(\rm{D}_2) \in [2.70, 3]$. Then the mean values are $\beta(\rm{H}_2)=2.89$ and $\beta(\rm{D}_2)=2.85$. Thence $\beta(\rm{H}_2) \simeq \beta(\rm{D}_2) \simeq 2.87$ is compatible with our assumption stipulating a common value of $\beta$ (the value obtained with our angular distribution was $\beta=2.5$). With this value $\beta=2.87$, we obtain the sticking coefficients $S_{\rm{H}_2}=0.12$ and $S_{\rm{D}_2}=0.28$: those are now in good agreement with the data of Govers {\it et al.}.
 
 We conclude that these different comparisons confirm both our physical model for the sticking process, and the description of the beam angular effect in these experiments (though our angle-dependent velocity distribution is not universal). Moreover this analysis shows that these different data must not be directly compared, and this explains the apparent discrepancy between them. This analysis implies also that these experimental sticking coefficients are different from the true thermal coefficients (probability of sticking with a gas phase at thermal equilibrium) suitable for astrophysical applications.

\begin{table}
\begin{ruledtabular}
\begin{tabular}{cccccc}
&X$_2$ &$\theta$ &$T$ (K)&$S$ (measure)&$S$ (model)\\
\hline
Govers & H$_2$ & $45^\circ$ & $293$ & $0.08 \pm 0.05$ & $0.18$ \\
Govers & D$_2$ & $45^\circ$ & $293$ & $0.27 \pm 0.05$ &  $0.37$\\
Hornekaer & D$_2$ & $0^\circ$ & $300$ & $0.2 \pm 0.15$ &  $0.25$\\
Amiaud & D$_2$ & $43^\circ$ & $293$ & $0.38 \pm 0.05$ &  $0.36$\\
This study & H$_2$ & $62^\circ$ & $300$ & $0.28 \pm 0.06$ & $0.24$\\
This study & D$_2$ & $62^\circ$ & $299$ & $0.42 \pm 0.06$ & $0.43$\\
\end{tabular}
\end{ruledtabular}
\caption{\label{tab:table1} Comparison between the sticking coefficients obtained in different experiments at room temperature (Govers {\it et al.},\cite{govers1980} Hornekaer {\it et al.},\cite{Hornekaer2003} Amiaud {\it et al.},\cite{amiaud2007} our results) and our model.}
\end{table}

\section{Discussion}
We are perfectly aware that the detailed interaction (and behavior) of atoms and molecules with the ASW ice surface are not similar. Moreover in each situation (atomic or molecular), we omit a lot of parameters. However one of the main interests of this kind of simple statistical model is precisely to erase the details and to keep the main features. The model could have been more complete if we were able to deduce (or compare) the velocities $c_0$ directly from other results. 

Choosing the dependence of sticking on velocity rather than on kinetic energy can be justified by the following arguments. On one hand the gas-surface interaction is based on short-ranged potentials that implies a finite distance $L$ of interaction, and on the other hand sticking implies an energy transfer to the surface that needs a minimal time $\tau_0$ of interaction. If the impinging particle has a velocity $v$, it interacts with the surface during a time $\tau \simeq 2 L/v$.  Then the particle sticks to the surface if $\tau > \tau_0$ that is $v < c_0=2L/\tau_0$. Then there exists a characteristic velocity $c_0=2L/\tau_0$ relative to sticking.

From the fits achieved for the atomic and molecular species (Appendix A and Sec. \ref{subsec:CompwithExp}), we can compare the velocity parameters $c_0(H)$ and $c_0(H_2)$. Since atomic and molecular hydrogen have not the same electronic structure, we expect  $c_0(H_2) \ne c_0(H)$. From the values $m_H c_0^2(H)=2 E_0=104 \text{ K}$  (Appendix \ref{subsec:TestwithBuch}) and $0.5 m_{H_2} c_0^2(H_2)= m_H c_0^2(H_2)=T_0(H_2)=87\text{ K}$ (Sec. \ref{subsec:CompwithExp}) we obtain
\begin{equation}
\frac{c_0(H_2)}{c_0(H)}  \simeq 0.91.
\end{equation}

As expected $c_0(H_2) \ne c_0(H)$, but the values are close. If we use our estimate $c_0$=2~$L$/$\tau_0$ where $L$ is a maximal distance of interaction and $\tau_0$ a minimal time of interaction, and if we assume roughly that $\tau_0(H_2)~\simeq~\tau_0(H)$ (we have not reached a regime of limited energy transfer), we find
\begin{equation}
\frac{L(H_2)}{L(H)} \simeq 0.91.
\end{equation}

This result is compatible with the idea that the range of H$_2$-surface interaction is shorter than that of the H-surface interaction: it is more difficult to trap H$_2$ than to trap H in a potential well on the surface, because H$_2$ has internal motions with possible energy transfers between the different degrees of freedom of the molecule (rotation-translation of the two atoms). Thus if the two atoms of the molecule approach the critical interaction distance $L(H)$ where only one H atom is trapped, this does not guarantee that the molecule will be trapped in the potential well and might be kicked out from the surface. This situation cannot take place in the case of only one H atom impinging on the surface.

The angular dependence of the velocity distribution seen by the sample, and then of the sticking probabilities measured (for experiments based on effusive beams), is a key point to understand the discrepancy between the different experimental values. As a consequence, these values are not the ``thermal" sticking coefficient (probability of sticking with a gas phase at thermal equilibrium). Astrophysicists can be interested only in this coefficient.
It can be deduced from Eqs. \ref{eqn:expdistrib}, \ref{sdshformula} and \ref{eqn:phicomplete}, by remarking that the thermal velocity modulus distribution (Boltzmann law) takes the form $v^2$ multiplied by the Boltzmann exponential factor. So this corresponds formally to the case $\alpha(\theta)=2$ of our angle-dependent velocity distribution (Eq.\ref{eqn:expdistrib}). We then deduce $\beta(\theta)=2.5$ from Eq. \ref{eqn:phicomplete}. The equation \ref{sdshformula} gives the final result. Then the true thermal coefficient is given (with our notations) by
\begin{equation}
\label{eqn:thermalstick}
S(T)=S_0 \frac{1+ \beta T/T_0}{(1+ T/T_0)^\beta},
\end{equation}
with $\beta=2.5$.

Another point can be analyzed from our model: it concerns the sticking probability of the molecule HD. It is reasonable to extend the mass law obtained for H$_2$ and D$_2$ to the case of HD. We deduce that the temperature parameter $T_0(HD)$ verifies $T_0(HD)=(3/2) T_0(\rm{H}_2)=130.5 \rm{K}$. Moreover since $S_0(\rm{H}_2)=0.76$ and $S_0(\rm{D}_2)=0.83$, we can estimate $S_0(\rm{HD}) \simeq 0.8$. With these parameters we obtain the thermal HD sticking coefficient from Eq. \ref{eqn:thermalstick}.

The values of the parameters $S_0$ and $T_0$ obtained for H$_2$ and D$_2$ (from our experiments: Sec. \ref{subsec:CompwithExp}), for H and D (from V. Buch computations: Appendix A) and for HD (prediction) are summarized in the table \ref{tab:table2}. 

\begin{table}
\begin{ruledtabular}
\begin{tabular}{ccc}
&$S_0$ &$T_0$(K) \\
\hline
H & $1$ & $52$\\
D & $1$ & $104$\\
H$_2$ & $0.76$ & 87\\
D$_2$ & $0.83$ & 174\\
HD & 0.8 & 130.5\\
\end{tabular}
\end{ruledtabular}
\caption{\label{tab:table2} Table of the different coefficients obtained in this article.}
\end{table}

\section{Conclusion}

In this paper we have reported a set of experiments that we have conducted to measure the sticking coefficients of molecular hydrogen and deuterium on $np$-ASW ice surfaces held at 10~K using the King \& Wells method. A study of the variation of the sticking coefficients with the molecular beam temperature was also presented. To our knowledge, this is the first experimental work that measures the sticking coefficient of hydrogen and deuterium molecules as a function of the beam temperature.

We have also presented an original model that explains the isotopic effect and the temperature behavior of the obtained experimental data. The model succeeds in fitting the present data for molecular species (H$_2$,D$_2$) as well as previous experimental results on (H$_2$,D$_2$) and theoretical ones of Buch \& Zhang on atomic species (H,D). To our knowledge, it is the first time that a simple model reproducing a so wide range of results is proposed. 

Following our model, we determined the values of the physical parameters describing the sticking process independently of any specific experimental situation ($S_0$, $T_0$, function $\Phi$). Then from these values, we have been able to propose an explicit formula giving the true thermal sticking coefficient (Eq. \ref{eqn:thermalstick}) for H$_2$, D$_2$, H and D.

The astrophysical interest of this article is that we can now extract some useful values of the thermal sticking coefficients of H$_2$ and D$_2$ (Eq. \ref{eqn:thermalstick}, Tab. II) relevant to the dark clouds of the interstellar medium. 

\begin{acknowledgments}
We acknowledge the support of the national PCMI program founded by the CNRS, as well as the strong financial support from the Conseil Regional d'Ile de France through SESAME programs (E1315 and I-07-597R) and the Conseil G\'en\'eral du Val d'Oise. We are grateful to the Agence Nationale de la Recherche (ANR) that supports this work in the framework of IRHONI Contract (No. ANR-07-BLAN-0129-2). One of us wants to thank V. Sidis for his help in the manuscript typesetting.
\end{acknowledgments}

\appendix

\section{\label{subsec:TestwithBuch} Testing our model on atomic sticking probabilities of H and D}

\begin{figure}
\resizebox{\hsize}{!}{\includegraphics{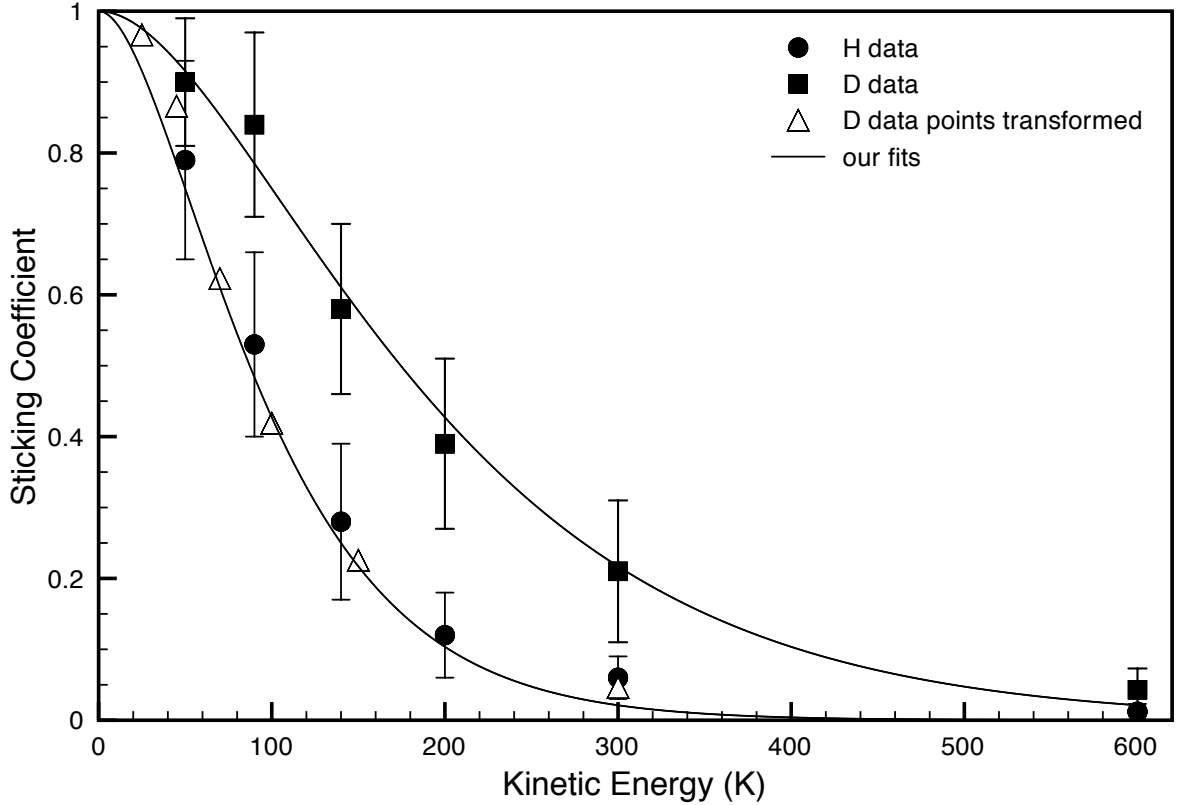}}
\caption{Sticking of H and D from Ref. \onlinecite{buch1991}. Closed circle for H, closed square for D. Open triangle: D data points transformed with Eq. \ref{probasticke}, ($S_0(H)/S_0(D)=0.95$). Fit of H and D data points based on Eq. \ref{eqn:stickbuch}.  $S_0(H)=S_0(D)=1$ and $E_0=52 \text{ K}$.} \label{figbuchzhang}
\end{figure}

In their theoretical article on the sticking probabilities of H and D atoms on clusters of amorphous water ice (cluster temperature $T \simeq 10 \text{ K}$ similar to our surface temperature), Buch \& Zhang\cite{buch1991} give a set of values for $S_H(E)$ and $S_D(E)$. These quantities correspond to our coefficient $S(G,v)$ because $E = (1/2) m v^2$. Moreover the function $\Phi$ of our model only depends on water ice properties and then it must be roughly the same. These data allow us to get rid of the average effect on the sticking coefficients due to our experimental velocity distribution and we can test directly the expression of $S(G,v)$ and our function $\Phi$. From our model (Eq. \ref{probastickv}) we have
\begin{equation}
S(E)=S_0 \Phi \left( \sqrt{\frac{2E}{m c_0^2}} \right).
\end{equation}
Let us notice that the velocity parameter $c_0$ involved in this equation is not necessarily the same as the one previously defined for the molecular case, because atomic and molecular hydrogen have not the same electronic properties. Using the procedure of Sec. \ref{subsec:MassandTemp}, we obtain the scaling law (renormalization-dilation transform)
\begin{equation}
S_{H}(E/2)=\frac{S_{0}(H)}{S_{0}(D)}S_{D}(E),
\label{probasticke}
\end{equation}
where the factor $2$ is the mass ratio between D and H. The effect of Eq. \ref{probasticke} on their data can be tested independently of any fit as in Sec. \ref{subsec:CompwithExp}. This is shown on Fig. \ref{figbuchzhang}.

Using our function $\Phi(x)=(1+x^2)e^{-x^2}$, we have
\begin{equation}
\label{eqn:stickbuch}
\left\{
\begin{array}{l}
S_{H}(E)=S_{0}(H)F(\frac{E}{E_0})\\
S_{D}(E)=S_{0}(D)F(\frac{E}{2E_0})
\end{array},
\right.
\end{equation}
where $F(x)=(1+x)e^{-x}$ and $E_0=E_0(H)=(1/2) m_H c_0^2$. The corresponding fits for atomic hydrogen and deuterium data are given on Fig. \ref{figbuchzhang}. They confirm that our choice of the function $\Phi$ is very satisfactory. Moreover this test, made on a complete different kind of data, shows the strength of our model and confirms the validity of our general physical assumptions.

\end{document}